# Automated Activity Recognition of Construction Equipment Using a Data Fusion Approach

Behnam Sherafat[1], Abbas Rashidi[2], Yong-Cheol Lee[3], Changbum R. Ahn[4]

[1]Behnam Sherafat, Ph.D. Student of Construction Engineering, Department of Civil and Environemental Engineering, University of Utah, UT, USA; email: behnam.sherafat@utah.edu
[2]Abbas Rashidi, Assistant Professor, Department of Civil and Environmental Engineering University of Utah, UT, USA; email: abbas.rashidi@utah.edu
[3]Yong-Cheol Lee, Assistant Professor, Department of Construction Management, Louisiana State University, LA, USA; email: yclee@lsu.edu
[4]Changbum R. Ahn, Associate Professor, Department of Construction Science, Texas A&M University, TX, USA; email: ryanahn@tamu.edu

## ABSTRACT

Automated monitoring of construction operations, especially operations of equipment and machines, is an essential step toward cost-estimating, and planning of construction projects. In recent years, a number of methods were suggested for recognizing activities of construction equipment. These methods are based on processing single types of data (audio, visual, or kinematic data). Considering the complexity of construction jobsites, using one source of data is not reliable enough to cover all conditions and scenarios. To address the issue, we utilized a data fusion approach: This approach is based on collecting audio and kinematic data, and includes the following steps: 1) recording audio and kinematic data generated by machines, 2) preprocessing data, 3) extracting time- and frequency-domain-features, 4) feature-fusion, and 5) categorizing activities using a machine-learning algorithm. The proposed approach was implemented on multiple machines and the experiments show that it is possible to get up to 25% more-accurate results compared to cases of using single-data-sources.

## INTRODUCTION

In the construction industry, it is well known that the ownership, rental, and maintenance costs of heavy equipment would significantly impact the overall budget and schedule of projects (Cheng et al. 2017, Ahn et al. 2012). As a result, it is vital to continuously monitor various operations of those equipment under different conditions. Manually recognizing those activities are time-consuming and costly, so the researchers and practitioners have suggested automated techniques for handling this important task. Results of an automated activity recognition system could be further used to identify idle times, calculate productivity rates, and estimate cycle times for repetitive activities (Rezazadeh and McCabe 2011, Teizer et al. 2010). Within the last couple of years, several studies have been conducted on this topic. Three major methods have been utilized for automated activity detection in construction job sites: 1) computer vison-based methods (Yang et al. 2016, Golparvar-Fard et al. 2013); 2) kinematic-based methods (Akhavian and Behzadan 2012, Ahn et al. 2013, Kim et al. 2018); and 3) audio-based methods (Zhang et al. 2018, Cheng et al. 2017a, Sabillon et al. 2017, Cheng et al. 2017b). Because of the scope of this paper, recent studies on audio- and kinematic-based studies are investigated in more detail within the following paragraphs.

Akhavian and Behzadan (2012) proposed a method that is capable of detecting the motion of different parts of the equipment using magnetic field and tilt sensing, and creating 3D simulations. This method can provide simulation models with accurate input data. In order to improve their framework, Akhavian and Behzadan (2013) and Akhavian and Behzadan (2014) used data fusion by merging weight, position, and orientation to detect equipment activities. Ahn et al. (2013) proposed a method to detect engine-off, idling, and working activities of excavators using accelerometer data. Recently, Kim et al. (2018) analyzed the cabin rotation data, which are recorded by Inertial Measurement Units (IMUs), by using a dynamic time wrapping technique to detect activities of excavators. The output of their method can determine the activity cycle time of the excavator.

Furthermore, a few studies have been conducted using audio data. Recently, Zhang et al. (2018) identified six types of equipment sounds using Mel-Frequency Cepstral Coefficients (MFCCs) and Hidden Markov Model (HMM) and obtained an accuracy of 94%. Cheng et al. (2017a), Cheng et al. (2017b), and Sabillon et al. (2017) implemented activity detection methods based on the audio recorded while the equipment operates. These methods use Short Time Fourier Transform (STFT) of audio signals to extract frequency-domain features.

All of the aforementioned studies are widely used either kinematic or audio data but have their specific issues. For example, audio might not work efficiently on congested, noisy and large job sites. Because the energy of audio diminishes in longer distances. Also, there are some issues with the kinematic-based methods. For example, some types of equipment (e.g. hand drills) do not provide enough space to place the sensors. To address these limitations, we propose an automated construction activity detection method using both audio and kinematic sensors simultaneously. The authors evaluated the applicability of this method on different job sites with several types of equipment such as excavators, loaders, and trucks from which they obtained very promising results.

**RESEARCH METHODOLOGY**

This paper proposes an automated method to detect various types of activities performed by construction equipment using audio and kinematic signals. To achieve this goal, the authors developed a binary classification to detect major (value-adding) and minor (non-value-adding) activities. Value-adding activities (e.g. moving soil) are those activities directly related to the productivity of the activity. On the other hand, non-value-adding activities (e.g. rotating cabin, maneuvering) are the ones that support the performance of value-adding activities and do not impact on the final productivity. Different steps of the proposed system are shown in Fig. 1.

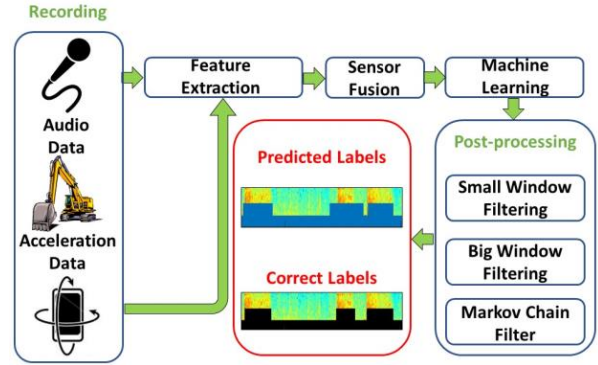

**Fig. 1. The proposed hybrid kinematic-acoustic system for activity detection of construction equipment**

**RECORDING DATA**

In this research, two types of data are recorded. The first one is audio, which is recorded using a single microphone placed outside the equipment cabin. The other type of data is kinematic, which is recorded using acceleration and angular velocity sensors. Three axes (x, y, and z) for each data type is recorded. Sampling frequencies for audio and kinematic sensors are 44100 Hz and 100 Hz, respectively.

**FEATURE EXTRACTION**

In this step, different features are extracted in the time and frequency domain. Segments of 120ms have been chosen for feature selection. This segment duration has been approved by previous studies for providing enough time resolution. Short Time Fourier Transform (STFT) with a Hanning window (50% overlap) is selected for feature extraction. Features are chosen based on their performance in previous studies (Table 1). Spectral features are shown to have good performance on detecting equipment activities based on the sound of equipment engine. All of the features are extracted and fed into the machine learning model described in section 3.4.

**Table 1. List of extracted features for both audio and kinematic data**

| Features | Usage | Reference |
|---|---|---|
| 25 STFT Coefficients | Audio & Kinematic | (Cheng et al. 2017) |
| Root Mean Square (RMS) | Audio & Kinematic | (Akhavian and Behzadan 2015) |
| Short Time Energy (STE) | Audio & Kinematic | (Akhavian and Behzadan 2015) |
| Spectral Flux (SF) | Audio & Kinematic | (Kozhisseri and Bikdash 2009, Padmavathi et al. 2010) |
| Spectral Entropy (SE) | Audio & Kinematic | (Wieczorkowska et al. 2018, Padmavathi et al. 2010) |
| Spectral Centroid (SC) | Audio & Kinematic | (Wieczorkowska et al. 2018, Padmavathi et al. 2010) |
| Spectral Roll-off (SRO) | Audio & Kinematic | (Wieczorkowska et al. 2018, Padmavathi et al. 2010) |
| Zero Crossing Rate (ZCR) | Kinematic | (Wieczorkowska et al. 2018) |

**SENSOR FUSION**

All of the features explained in the previous section are extracted from the audio and kinematic signals. As the next step, the authors used a feature fusion approach to combine features from both audio and kinematic data to achieve a more robust system. Fused features are used as input to the machine learning model.

**SVM MODEL**

The authors implemented Support Vector Machine (SVM) as the machine learning algorithm. This algorithm has shown good performance in similar previous efforts (Sabillon et al. 2017). In this paper, different activities of equipment are recorded using a video camera for further actual labeling of activities. Four time periods of major activities and four time periods of minor activities with durations of 5 to 10 seconds have been considered as samples for training the model. The other time periods are used as testing data. SVM predicts the labels and these labels are then fed into the post-processing algorithms.

**POST-PROCESSING DATA**

The labels from the previous section are not appropriate for predicting activities. These labels fluctuate in very little periods of time, which is not practical for prediction purposes. In this research, three types of algorithms are used to smooth the labels: 1) Small Window Filtering (SWF); 2) Big Window Filtering (BWF); and 3) Markov Chain Filter (MCF). These algorithms look through the labels and substitute the labels with the more frequent label of their previous or future ones. Windows for checking the frequency of labels are chosen as 2 and 6 for small window SWF and BWF, respectively (Sabillon et al. 2017). More explanation about these three filters are provided in previous researches (Sabillon et al. 2017).

**EXPERIMENTAL SETUP AND RESULTS**

The proposed system is evaluated on 5 different types of equipment including a loader (CAT 259D), a dozer (Dozer 850K), an excavator (CAT 308E), a drilling machine (Jackhammer), and a lift (XTREME). The data capturing devices implemented in this study are Zoom H1 digital handy recorder, an off-the-shelf microphone, and iPhone for recording kinematic data. The microphone is placed outside the cabin within a distance of about 10ft and an iPhone is mounted inside the cabin. The iPhone and laptop are connected using MATLAB software to send data from the iPhone to the laptop. In addition, a video camcorder is used to record all activities of the equipment for further labeling and validating purposes. Fig. 2 shows the configuration of the devices.

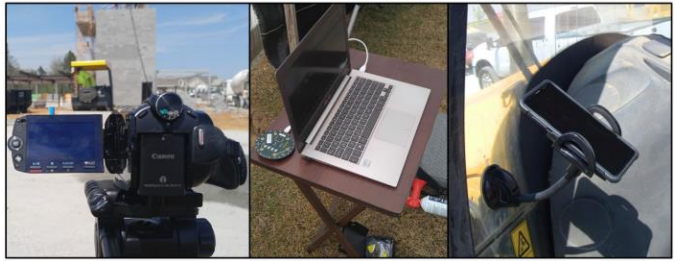

**Fig. 2. Placement of devices near and inside the cabin of the equipment**

Figure 3 demonstrates the generated signals after fusing data for the jackhammer. As shown in this figure, there is a strong correlation between the predicted and actual labels. Finally, the summary of the accuracies obtained for all five pieces of equipment is presented in Figure 4.

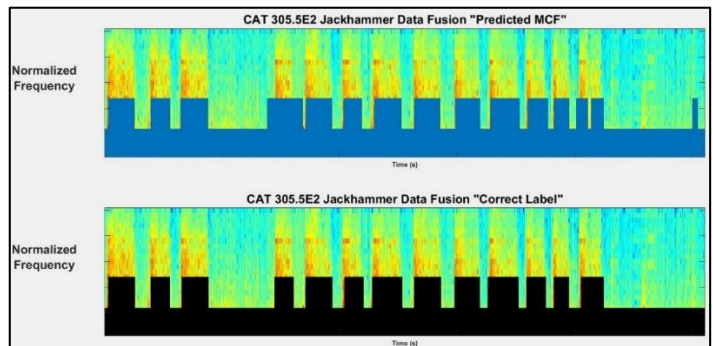

**Fig. 3. Predicted labels after MCF and actual labels for Jackhammer using fused data**

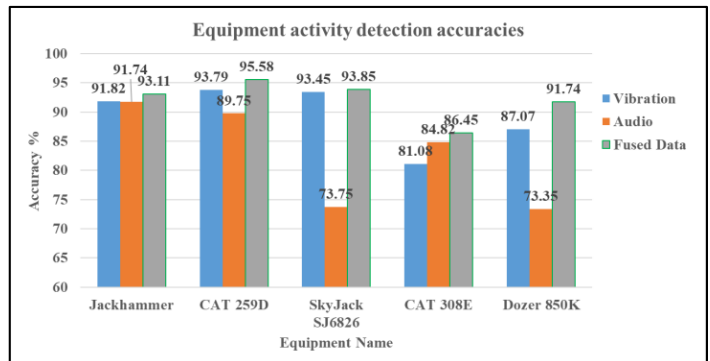

**Fig. 4. Equipment activity detection accuracies using two types of data and the fused data**

## DISCUSSIONS

This paper proposed a data fusion approach, based on using audio and kinematic data, for automated activity detection of construction equipment. While the proposed system shows very promising results, a couple of issues and challenges still exist: First, in some cases, it might be difficult to specify an activity as productive or non-productive. In other words, it depends on the judgement of project managers. For example, when a drill is attached to the arm of CAT 305.5E2 to break concrete, moving back and forth, the rotating and extending arm are considered as minor activities because they are not related to breaking concrete. Furthermore, the authors found that some activities could not be detected correctly. This is due to the fact that some types of equipment generate similar sounds when doing different tasks. For example, moving the arm forward and backward are major and minor activities, respectively, but they have similar sound patterns. For a Dozer 850K, some portions of moving backwards are incorrectly detected as a major activity.

For jackhammer, both data sources are appropriate for activity detection. This is due to the fact that drilling operations generate strong and distinct vibration and sound patterns. However, when it comes to smaller backhoes, the generated kinematic data are not strong enough, which leads to less accurate results. By observing the accuracies of dozer (i.e. Dozer 850K), it is almost clear that vibration is a better data source for activity detection. For excavator (CAT 308E), both data sources lead to same results. Because activities of excavators generate very distinct vibration and sound patterns, it is possible to detect activities accurately. The other possible reason for inaccuracies is the issue of time synchronization. The starting point of audio and kinematic signals might not be exactly the same, which might impact the accuracy of results negatively. In this study, this issue has been addressed using manual modifications. Also, some old types of equipment generate outlier kinematic signals which distort the desired kinematic signals and decrease the accuracy. On the other hand, some newer models are quieter and tends to generate lower levels of noise and vibrations, which could lead to lower accuracy rates. A detailed analysis of the results shows that each of audio and kinematic sensors works better with certain types of equipment, and might not be appropriate for other types.

In conclusion, evaluating the results shows that audio sensors are a good choice for excavators and kinematic sensors work better for bulldozers. For loaders, both types of data lead to similar results. In Table 2, a brief comparison between different types of activities and sensors is presented.

**Table 2. Accuracy rates for kinematic data, audio data, and the fused data**

| Equipment | Vibration Accuracy | | | Audio Accuracy | | | Fused Data Accuracy | | |
|---|---|---|---|---|---|---|---|---|---|
| | Low (0%-75%) | Moderate (75% - 90%) | High (90%-100%) | Low (0%-75%) | Moderate (75% - 90%) | High (90%-100%) | Low (0%-75%) | Moderate (75% - 90%) | High (90%-100%) |
| Jackhammer | | | ✓ | | | ✓ | | | ✓ |
| CAT 259D | | | ✓ | | ✓ | | | | ✓ |
| Skyjack SJ6826 | | | ✓ | ✓ | | | | | ✓ |
| CAT 308E | ✓ | | | | ✓ | | | | ✓ |
| Dozer 850K | ✓ | | | ✓ | | | | | ✓ |

In summary, fusing both audio and kinematic signals helps detect equipment activities in a more robust and accurate way and the results show that fusing data leads to accuracy rates over 90%. In the next section, some other advantages of the system as well as future research plans are explained.

**CONCLUSION**

The proposed system automatically detects different types of activities performed by construction equipment using the generated audio and kinematic signals. This system consists of recording data, preprocessing data, extracting several features, feature fusion, and classifying activities using the SVM model. The contributions of the proposed system are as follows: 1) it can address the issue of using a single type of sensor. For example, in congested and/or large job sites, the energy of audio might decrease over large distances. Also, some new models do not generate detectable kinematic signals, which decreases the accuracy of the system; 2) in this paper, different time and frequency-domain features are extracted, which support the process of machine learning model. All of these features are evaluated in previous studies and showed accurate results in detecting equipment engine sound; and 3) different pre-processing and post-processing algorithms are conducted to refine the data and results.

This paper focused on detecting the activities of single pieces of equipment. In the future, the authors will evaluate this method on more challenging cases where multiple machines operate at a jobsite simultaneously. In addition, there is a need to collect more data from other types of equipment such as graders and compactors.

## ACKNOWLEDGMENTS

This research project has been funded by the U.S. National Science Foundation (NSF) under Grant CMMI-1606034. The authors gratefully acknowledge NSF's support. Any opinions, findings, conclusions, and recommendations expressed in this manuscript are those of the authors and do not reflect the views of the funding agency. The authors also appreciate the assistance of Mr. Richard Peterson, undergraduate student at University of Utah, with data collection and audio recordings.